\documentclass[12pt]{article}
\usepackage{amsmath, amssymb}
\begin{document}

\title{
MASS GAP\\
IN\\
QUANTUM CHROMODYNAMICS}

\author{R. Acharya \footnote{Dedicated to honor George Sudarshan on his $75^{th}$ birthday}\\
Department of Physics \& Astronomy\\
Arizona State University\\
Tempe, AZ 85287-1504}

\date{November 2006}
\maketitle

\begin{abstract}
We present a heuristic argument in support of the assertion that
QCD will exhibit a mass gap, if the Callan-Symanzik function
$\beta(g)$ obeys the inequality $\beta(g) < 0$, for
\underline{all} $g > 0$.
\end{abstract}

We begin by summarizing the standard lore attributed to QCD: (A)
QCD must have a mass gap, i.e., every excitation above the vacuum
state must have energy $\geqslant \bigtriangleup$, where
$\bigtriangleup$ is a positive constant. The key idea that
infrared slavery requires the generation of a mass gap (dynamical
gluon mass) in QCD was enunciated in a pioneering paper by Cornwall
in 1982 \cite{Cornwall}. (B) QCD must exhibit ``quark
confinement", i.e., the observed (physical) particle spectrum
(pion, proton,...) are color $SU_{C}(3)$ invariant, despite the
fact that QCD is described by an underlying Lagrangian of quarks
and non-abelian gluons, which transform non-trivially under color
$SU_{c}(3)$ symmetry. (C) QCD must exhibit chiral symmetry
breakdown, i.e., the flavor axial-vector charges $Q^{\alpha}_{5}
(\alpha = 1, 2 \ldots 8)$ must break the flavor $SU^{L}_{F} (3)
\otimes SU^{R}_{F} (3)$ spontaneously, i.e. $Q^{\alpha}_{5} \quad |_{0} \rangle
 \neq 0$, so that the vacuum is only invariant under a
subgroup of the full symmetry group acting on the quark fields, in
the limit of vanishing current quark masses (Number of flavors,
$F=6$) \cite{Weinberg}.

Item (A) is essential to understand the short-range of the nuclear
force. Item (B) is essential to account for the absence
(unobservability) of individual free quarks. Finally, item (C) is
essential to justify the spectacular current algebra predictions
of the 1960's.

In this note, we present a heuristic argument to validate Item I,
under the following assumptions:

\begin{itemize}

\item[(a)] Elitzur's theorem holds \cite{Elitzur}: local gauge invariance
cannot be spontaneously broken: local gauge invariance is really a
tautology \cite{Rajagopal}, stating the redundancy of variables.
As a consequence, the vacuum is generically non-degenerate and
points in ``no particular direction" in group space and there
cannot be massless, Nambu-Goldstone bosons (Colored, in QCD)
\cite{Englert}.

\item[(b)] Federbush-Johnson-Schroer theorem holds \cite{Federbush}: In a
Lorentz-invariant quantum field theory, if a local operator
annihilates the vacuum state, then the operator must vanish
identically. We will elaborate on this theorem later, with a
crucial and important clarification due to Greenberg
\cite{Greenberg}.

\end{itemize}

We now sketch the argument leading to the ``proof" of Item I,
i.e., the existence of a mass gap $\bigtriangleup > 0$, in QCD.

We begin with the conserved, vector (color)current $V^{a}_{\mu}
(a=1,2 \ldots 8)$:

\begin{equation}\label{eq1}
\partial^{\mu} V^{a}_{\mu} (\overrightarrow{x}, t) = 0
\end{equation}

This implies the validity of the \underline{local} version, i.e.,

\begin{equation}\label{eq2}
\left[ Q^{a}, \quad H (\overrightarrow{x}, t) \right] = 0
\end{equation}
where $H (\overrightarrow{x}, t) = \ominus^{00}$ is the
Hamiltonian density of QCD, if the surface terms at infinity can
be discarded. This is clearly justified and guaranteed by
Elitzur's theorem \cite{Elitzur} which insures that

\begin{equation}\label{eq3}
Q^{a} \quad |_{0} \rangle = 0 \quad , Q^{a} = \int d^{3}_{x} V^{a}_{0}(x)
\end{equation}

Since local color $SU_{C} (3)$ ivariance cannot be spontaneously
broken and hence there \underline{cannot} be massless, colored
Nambu-Goldstone bosons: the QCD vacuum is generically
non-degenerate and points in ``no particular direction" in group
space \cite{Englert}. We note in passing that Eq.(\ref{eq2}) is
\underline{stronger} than the consequence of Coleman's theorem
\cite{Coleman}:

\begin{equation}\label{eq4}
Q^{a} \quad |_{0} \rangle = 0 \Rightarrow [Q^{a},H] = 0.
\end{equation}

Eq.(\ref{eq2}) is a \underline{local} version of Eq.(\ref{eq4}).

The dilatation charge $Q_{D}(t)$ is defined by

\begin{equation}\label{eq5}
Q_{D}(t) = \int d^{3}_{x} D_{0} (\overrightarrow{x}, t)
\end{equation}
where $D_{\mu}$ is the dilatation current whose divergence is
determined in QCD by the trace anomaly \cite{Adler}:

\begin{equation}\label{eq6}
\partial^{\mu} D_{\mu} = \frac{\beta(g)}{2g} F^{a}_{\mu\nu}
F^{\mu\nu ^{a}}
\end{equation}
in the limit of vanishing current quark masses, $m_{i} = 0$, $i =
u,d,s$.

Eq.(\ref{eq6}) expresses the explicit breakdown of scale
invariance, via the appearance of $\beta (g)$.

The scale dimension $d_{Q}$ is defined via the commutator relation
\cite{Acharya},

\begin{equation}\label{eq7}
[ Q_{D} (0), \quad Q^{a}(0) ] = - i d_{Q} Q^{a} (0)
\end{equation}

Since the vector current $V^{a}_{\mu}$ is conserved
(Eq.(\ref{eq1})), the associated charge $Q^{a}$ has zero scale
dimension, ($d_{Q} = 0$).

Eq.(\ref{eq7}) can be ``promoted" to read

\begin{equation}\label{eq8}
[Q_{D} (t), \quad Q^{a} ] = 0
\end{equation}
by time translation, keeping in mind the validity of
Eq.(\ref{eq4}).

Consider the double-commutator, $[Q_{D} (t), \quad [Q^{a}, H
(\overrightarrow{x}, t)] ]$ which can be manipulated with the help
of Jacobi's identity, to yield:
\begin{equation}\label{eq9}
\begin{split}
& \left[ Q_{D} (t), \quad \left[ Q^{a}, H(\overrightarrow{x}, t) \right]
\right] \\
& = - \left[ Q^{a}, \left[ H(\overrightarrow{x}, t),  Q_{D} (t) \right]
\right] \\
& - \left[ H (\overrightarrow{x}, t), \quad \left[ Q_{D} (t), Q^{a} \right] \right]
\end{split}
\end{equation}

This simplifies to, in view of Eq.(\ref{eq2}) and Eq.(\ref{eq8}):

\begin{equation}\label{eq10}
\left[ Q^{a}, \quad \left[ H(\overrightarrow{x}, t), Q_{D} (t)
\right] \right] = 0
\end{equation}

The trace anomaly, Eq.(\ref{eq6}) leads Eq.(\ref{eq10}) to the
result:

\begin{equation}\label{eq11}
\left[ Q^{a}, \quad \partial^{\mu} D_{\mu} \right] = 0
\end{equation}

Since,
\begin{equation}\label{eq12}
i \left[H(\overrightarrow{x}, t), Q_{D} (t) \right] = \partial^{\mu}
D_{\mu}\neq 0
\end{equation}

Eq.({\ref{eq11}) leads to:

\begin{equation}\label{eq13}
\left[ Q^{a}, \partial^{\mu} D_{\mu} \right] \quad |_{0} \rangle = 0
\end{equation}

In view of Elitzur's theorem \cite{Elitzur}, Eq.(\ref{eq3}), we
arrive at the result

\begin{equation}\label{eq14}
Q^{a} \partial^{\mu} D_{\mu} \quad |_{0} \rangle = 0
\end{equation}

At this juncture, we follow Greenberg \cite{Greenberg} to
elucidate the ``meaning" of locality in the context of a quantum
field theory. As Greenberg has emphasized that ``the property of
locality can have \underline{three} different meanings
\cite{Greenberg} for a quantum field theory, (i) the fields enter
terms in the Hamiltonian and the Lagrangian at the
\underline{same} spacetime point, (ii) the observables commute at
space-like separation, and (iii) the fields commute (for integer
spin fields) or anticommute (for odd half-integer spin fields) at
space-like separation."

Greenberg continues to state the following \cite{Greenberg}:
``Theories in which (i) fails can still obey (ii) and (iii), for
example, quantum electrodynamics in Coulomb gauge. Theories in
which (iii) fails can still obey (i) and (ii); for example,
parastatistics of order greater than one; the theory in which CPT
is violated due to having different masses for the particles and
antiparticles is nonlocal in sense (iii); such a theory will be
nonlocal in sense (ii) and in sense (i)."

We now address the issue of locality with reference to
Eq.(\ref{eq14}) and we elaborate the point, as it plays a crucial
role in our subsequent analysis of Eq.(\ref{eq14}).

We postulate (the `` obvious"!) that the divergence of the scale
current $D_{\mu}$ is local in the sense (iii), outlined by
Greenberg:

\begin{equation}\label{eq15}
\left[ \partial^{\mu} D_{\mu} (x), \partial^{\mu} D_{\mu} (y)
\right]_{-} = 0, \quad x \sim y
\end{equation}

Since Eq.(\ref{eq11}), asserts that

\begin{equation*}\tag{11}
\left[ Q^{a}, \partial^{\mu} D_{\mu} \right] = 0
\end{equation*}

We conclude that

\begin{equation}\label{eq16}
\left[ Q^{a} \partial^{\mu} D_{\mu} (x), Q^{a} \partial^{\mu} D_{\mu}
(y) \right] = 0, \quad x \sim y
\end{equation}

Once again (!), we \underline{quote} (!) Greenberg
\cite{Greenberg}: ``Eq.(\ref{eq16}) seems to have nonlocality
because of the space integral in the $Q$ factors; however, if
[this is Eq.(\ref{eq15})

$$
\left[\partial^{\mu} D_{\mu} (x), \partial^{\mu} D_{\mu}
(y) \right]_{-} = 0, \quad x \sim y
$$
then \underline{Eq.(\ref{eq16}) holds, despite the apparent
nonlocality. What is relevant is} \\
\underline{the commutation relation, not the representation in
terms of a space} \\ \underline{integral."}

In other words, Eq.({\ref{eq16}) expresses the statement of
\underline{locality}, in the sense (iii), as outlined by Greenberg
\cite{Greenberg}.

We now invoke assumption (b), i.e., Federbush-Johnson-Schroer
Theorem \cite{Federbush} and arrive at the conclusion:

\begin{equation}\label{eq17}
Q^{a} \partial^{\mu} D_{\mu} \equiv 0
\end{equation}

Eq.{\ref{eq17}} has \underline{two} possible solutions:

\underline{Solution `A':}

$$Q^{a} \neq 0, \quad Q^{a} |_{0}\rangle = 0$$

\begin{equation}\label{eq18}
\Longrightarrow \partial^{\mu} D_{\mu} = 0
\end{equation}

\underline{Solution `B':}
\begin{equation}\label{eq19}
\partial^{\mu} D_{\mu} \neq 0 \Longrightarrow Q^{a} \equiv 0
\end{equation}

Solution `A' yields the Non-Abelian, Coulomb phase, corresponding
to an \underline{infrared} fixed point corresponding to a
nontrivial value of the coupling constant (i.e., neither zero
indicating triviality nor infinity, indicating ``confinement").
This means, that solution `A' yields a conformal field theory, in
the infrared: it has no mass gap and there are bound particles
whose mass is continuous and can take any positive value. This is
the Banks-Zaks fixed point (Infrared) \cite{Banks}.

\underline{Solution `B' is the option that concerns us here.}

In this case, the Callan-Symanzik function, $\beta(g)$ starts out
negative at the origin (asymptotic freedom!) and remains negative
as `$g$' increases and never turns over: there is \underline{No}
infrared fixed point at a nontrivial value of `$g$'. Consequently,
the non-abelian color charges $Q^{a} (a=1,2,\ldots,8)$ vanishes
identically!:

\begin{equation}\label{eq20}
Q^{a} \equiv 0
\end{equation}

i.e., the colored charges are Debye-screened \cite{Brydges}. This
has the implication that the non-Abelian gluons must acquire a
dynamical mass (i.e., mass gap) \underline{without} violating
local gauge invariance. This is a manifestation of the ``Higgs
(confinement) phase" \cite{Fradkin} and requires that all observed
(observable) particles must be color singlets, i.e., QCD must
exhibit ``quark confinement." This was the scenario envisaged by
Cornwall in 1982 \cite{Cornwall}.

\begin{center}
\textbf{Acknowledgement}
\end{center}

I wish to acknowledge the insightful conversation with Professor
'Hooft, on the outstanding unresolved questions in particle
theory, during his visit to Arizona State University.

\end{document}